\documentclass[openacc]{rstransa}

\usepackage{graphicx}
\usepackage{bbold}
\usepackage{epstopdf, epsfig}
\usepackage{amsmath,amssymb}
\usepackage{braket}
\usepackage[english]{babel}
\usepackage{ulem}
\usepackage{cancel}

\newcommand{\tu}{\tilde{ u}}

\newcommand{\bu}{{\bf u}} 
\newcommand{\bw}{{\bf w}} 
\newcommand{\bF}{{\bf f}}
\newcommand{\bk}{{\bf k}}
\newcommand{\bq}{{\bf q}}

\newcommand{\bx}{{\bf x}}

\newcommand{\bh}{{\bf h}}
\newcommand{\tbu}{\tilde{\bf u}} 
 
\newcommand{\tbF}{\tilde{\bf f}}

\newcommand{\cE}{{\mathcal{E}}}
\newcommand{\cH}{{\mathcal{H}}}

\newcommand{\cD}{{\mathcal{D}}}
\newcommand{\cC}{{\mathcal{C}}}
\newcommand{\Rep}{\mathrm{Re}_\epsilon}
\newcommand{\Reu}{\mathrm{Re}_{_U}}
\newcommand{\beq}{\begin{equation}}
\newcommand{\eeq}{\end{equation}}

\def\LB#1{\textcolor{black}{#1}}

\usepackage{textgreek}

\begin{document}

\title{$\lambda$-Navier-Stokes turbulence}

\author{
A. Alexakis$^{1}$ and L. Biferale$^{2}$}

\address{
$^{1}$ Laboratoire de Physique de l'Ecole normale sup{\'e}rieure, ENS, Universit{\'e} PSL, CNRS, Sorbonne Universit{\'e}, Universit{\'e} de Paris, F-75005 Paris, France \\ 
$^{2}$ Department of Physics and INFN University of Rome `Tor Vergata', Via della Ricerca Scientifica 1, 00133 Rome, Italy}

\subject{Fluid Dynamics, Statistical Mechanics}


\corres{A. Alexakis\\
\email{alexakis@phys.ens.fr}}

\begin{abstract}
We investigate numerically the model proposed in \cite{sahoo2017discontinuous} where a parameter $\lambda$ is introduced in the Navier-Stokes equations such that the weight of homochiral to heterochiral interactions is varied while preserving all original scaling symmetries and inviscid invariants. Decreasing the value of $\lambda$ leads to a change in the direction of the energy cascade at a critical value $\lambda_c \sim 0.3$. 
In this work, we perform numerical simulations at varying  $\lambda$ in the forward energy cascade range and at changing the Reynolds number $\mathrm{Re}$. We show that for a fixed injection rate, as $\lambda \to \lambda_c$, the kinetic energy diverges with a scaling law $\cE\propto (\lambda-\lambda_c)^{-2/3}$. 
The energy spectrum is shown to display a larger bottleneck as $\lambda$ is decreased.
The forward heterochiral flux and the inverse homochiral flux both increase in amplitude as $\lambda_c$ is approached
while keeping their difference fixed and equal to the injection rate. As a result, very close to $\lambda_c$ a stationary state
is reached where the two opposite fluxes are of much higher amplitude than the mean flux and large fluctuations are observed. 
Furthermore, we show that intermittency as $\lambda_c$ is approached is reduced.
The possibility of obtaining a statistical description of regular Navier-Stokes turbulence as an expansion around this newly found critical point is discussed. 

\end{abstract}


\begin{fmtext}
\section{Introduction}
In a turbulent flow energy is injected at  large scales  and dissipated at much smaller scales by viscosity \cite{frisch1995turbulence}. A transfer of energy is thus required from one scale to the other that is achieved by the energy cascade caused by the non-linearity of the Navier-Stokes equations.\end{fmtext}
\maketitle
\noindent
    The fundamental idea of the energy cascade across scales was first introduced by Richardson (1922) \cite{davidson2011voyage} and later quantified  by Kolmogorov \cite{kolmogorov1941local}. Under the assumption of scale-similarity, Kolmogorov predicted a power-law behaviour for the energy spectrum $E(k)\propto \epsilon^{2/3}k^{-5/3}$ and for the scaling of  the moments of velocity's differences across a distance $r$: $\langle |\delta_r u|^p\rangle \propto (\epsilon r)^{p/3}$.
 However, overwhelming experimental and numerical evidence have shown that the process of transferring energy from one scale to the other is not self-similar and that there exist anomalous exponents  such that   $\langle |\delta_r u|^p\rangle \propto (\epsilon r)^{\zeta_p}$ with $\zeta_p\ne p/3$. This is true for all moments
except for the third moment of the longitudinal velocity difference where $\zeta_3=1$ is exact. This departure from self-similarity has been 
phenomenologically explained in terms of multifractal theory and intermittency \cite{frisch1980fully}.   
There have been various attempts to predict and explain the observed exponents, see e.g. \cite{frisch1978simple,benzi1984multifractal,sheleveque,dubrulle1994intermittency}. However,  all these attempts  are based on simplified phenomenological assumptions and no exact or systematic derivation of the anomalous corrections directly from the Navier-Stokes equations has been proposed so far. 
{As a result, the existence of exactly solvable limits from where to develop perturbative or asymptotic expansions has been long sought.}
{
The main theoretical obstacle to attack three dimensional turbulent flows comes from being out-of-equilibrium, with anomalous scaling laws and  stronger and stronger non-Gaussian small-scales statistics at increasing Reynolds numbers. In general no universal recipes exist for the treatment of out-of-equilibrium problems. Equilibrium Gaussian, or quasi-Gaussian systems have predictable statistics but are only met in fluid dynamics for the 
truncated Euler equations where only a finite number of Fourier modes are kept. 
In this case, energy is conserved exactly and no finite energy flux through scales exists \cite{lee1952some,orszag1974lectures,kraichnan1973helical}. Although an expansion from such a state to a weakly cascading case can be performed \cite{alexakis2020energy} it seems unlikely to serve as a starting point to recover regular Navier-Stokes turbulence. Solvable
out-of-equilibrium states over which an expansion could be carried out have been sought with the use of re-normalization group theory.
Some of such studies consider deviations from a power-law force spectrum $F\propto k^{-d+(4- \epsilon)}$, where $\epsilon=d-2$ (with $d=3$ the dimension) corresponds to a molecular background noise, representing fluctuations in an equilibrium fluid at absolute temperature and $\epsilon=4$ corresponds to real turbulence 
\cite{forster1977large, fournier1983remarks, yakhot1986renormalization, eyink1994renormalization}. 
Other studies consider expansions from critical dimensions $d\gg 3$ where turbulence is conjectured to follow mean field dynamics, or from $d=4/3$ where the finite flux spectrum coincides with thermal spectrum, or at changing the couplings among triads in Fourier space \cite{kraichnan1961dynamics, forster1976long, siggia1977tricritical, bell1978time, fournier1978d, fournier1978infinite, frisch1976crossover, l2002quasi}. 
\LB{The effect of helicity has also been investigated with these techniques and found to play a minor role \cite{pouquet1978helicity,zhou1990effect,jurvcivsinova2009influence}. } 
Higher or non-integer dimensions however are not physically realisable and can be studied and tested with the use of numerical simulations only \cite{gotoh2007statistical, yamamoto2012local,frisch2012turbulence, lanotte2015turbulence, berera2020homogeneous}.}

{
More recently, systems where one dimension is compactified were demonstrated to result it a transition from three-dimensional behavior with a forward energy cascade, towards a two-dimensional behavior where energy cascades backward \cite{smith1996crossover,celani2010turbulence,benavides2017critical}. In this kind of transitions, the systems dimension $d$ does not vary continuously from $d=3$ to $d=2$ as in the previous considerations but have the advantage of being physically realizable.  Similar transitions have been observed in a variety of physical systems at varying some control parameters, including the rotation intensity, magnetic fields, stratification and forcing properties. For a recent review see \cite{alexakis2018cascades}.  In most of these cases the transition from forward to inverse cascade occurred through a split state where both fluxes exist simultaneously. Recently, a variant of the Navier-Stokes equations was introduced, with a dimensionless control parameter, $\lambda$, weighing the relative importance of homochiral and heterochiral triads and developing an abrupt transition from forward to inverse cascade at a critical value $\lambda_c$ \cite{sahoo2017discontinuous}. Right at the critical value a new singular state exist where energy doesn't cascade neither forward or inverse without the flow being necessarily at equilibrium.    }

{ Here, we study in greater detail  the behavior of the system when $\lambda$ is close but above this critical value. We argue that as the critical point is approached the flow is closer and closer to a flux-loop state where the mean energy flux towards the small scales is subdominant and large turbulent fluctuations that transfer energy both to large and small scales develop. We also demonstrate that intermittency in our model is reduced  as $\lambda \to \lambda_c$. This could indicate that the flux-loop state at $\lambda=\lambda_c$ has a more tractable statistics and could serve as starting point for perturbative expansion towards real Navier-Stokes turbulence at $\lambda=1$.}

\section{Formulation} 
\subsection{Helical decomposition} 
Let $\bu(t,\bx)$ be a zero-mean divergence-free vector field
defined in a cubic triple periodic domain of side $L$. 
Its Fourier transform $\tbu_\bk(t)$ is given by 
\beq
\tbu_\bk(t) = \frac{1}{L^3}\int \bu(t,\bx) \, e^{-i\bk \cdot \bx} dx^3
\quad\mathrm{and}\quad 
\bu(t,\bx) = \sum_{\bk} \tbu_\bk(t) e^{i \bk \cdot \bx}, 
\eeq 
where the three component complex vector $\tbu_\bk$ satisfies $\bk\cdot \tbu_\bk = 0$ due to the divergence-free condition. Thus, 
each $\tbu_\bk$ has two independent degrees of freedom. 
A convenient way to express these degrees of freedom is using the  helical mode 
decomposition \cite{craya1957contribution,herring1974approach,lesieur1972decomposition}, where $\tbu_\bk(t)$ is decomposed in two helical modes 
\beq 
\tbu_\bk(t) = \tu_\bk^+(t) \bh_\bk^+ +  \tu_\bk^-(t)(t) \bh_\bk^- .
\eeq 
Here, $\tu_\bk^\pm(t)$ are two independent complex scalar amplitudes
and the orthogonal unit vectors $\bh^\pm_\bk$ are given by:
\beq 
\bh^\pm_\bk = 
\frac{\bk \times \bk \times \hat{\bf e}}{\sqrt{2}k|\bk \times \hat{\bf e}|} \pm i 
\frac{\bk \times \hat{\bf e}}{\sqrt{2}|\bk \times \hat{\bf e}|} 
\eeq 
where $\hat{\bf e}$ is an arbitrary vector non-parallel to $\bk$.
The vectors $\bh^\pm_\bk$ are eigen-functions of the curl satisfying
$ 
i \bk \times \bh^\pm_\bk = \pm k \bh^\pm_\bk 
$ 
with $k=|\bk|$ and 
$\bh^{s_1}_\bk \cdot (\bh^{ s_2}_{ \bk})^*=
 \bh^{s_1}_\bk \cdot  \bh^{-s_2}_{ \bk}   =
 \bh^{s_1}_\bk \cdot  \bh^{ s_2}_{-\bk}   =\delta_{s_1,s_2}$ 
(where $s_i=\pm1$ and $\delta_{s_1,s_2}$ is the Kronecker delta). 
Using this decomposition we can split the real vector field $\bu(t,\bx)$ in two helical fields as  
$\bu(t,\bx)=\bu^+(t,\bx)+\bu^-(t,\bx)$ with $\bu^\pm(t,\bx)$ given by
\beq 
\bu^\pm(t,\bx) = \sum_\bk \tu^\pm_\bk(t) \bh^\pm_\bk e^{i\bk \cdot \bx} 
\eeq 

The velocity field $\bu(t,\bx)$ is evolved in time based on the Navier-Stokes equations (NSE)
\beq
\partial_t \bu + \bu \cdot \nabla \bu = -\nabla P + \nu \nabla^2 \bu +\bf f, 
\eeq 
where $P$ is the pressure enforcing incompressibility $\bf \nabla \cdot \bu=0$, $\nu$ is the viscosity and $\bf f$ is an external body force.
In particular, using the helical decomposition the Navier-Stokes can be written as
\beq
\partial_t \bu^{s_1} = \mathbb{P}^{s_1}\left[ \sum_{s_2,s_3} (\bu^{s_2}\times \bw^{s_3}) \right] + \nu \nabla^2 \bu^{s_1} + \bf f^{s_1}, 
\label{HNS}
\eeq 
where $s_i=\pm1$, $\bw^\pm = {\bf \nabla} \times \bu^\pm$, $\mathbb{P}^\pm$
is a projector to the helical base
\beq
\mathbb{P}^\pm\left[ {\bu(t,\bx)} \right] = \sum_\bk \bh^\mp_\bk \cdot \tbu_\bk(t) e^{i \bk \cdot \bx}   
\eeq 
and $\bF^{s_1}=\mathbb{P}^{s_1}\left[\bF\right]$. Note that the helical base is an incompressible base
and projecting in to it eliminates the pressure.
From the eight nonlinear terms that appear in eq.\ref{HNS}, one for each sign combination $s_1,s_2,s_3$,
the six that involve different signs are responsible for transferring energy to the small scales
while the two homochiral terms $s_1=s_2=s_3=\pm1$ transfer energy in the large scales \cite{waleffe_nature_1992,alexakis2017helically}. In the absence of forcing, viscosity (and in the absence of singularities) the evolution of $\bu(t,\bx)$ conserves two  ideal invariants the energy $\cE$ and the helicity $\cH$:
\beq 
\cE=\frac{1}{2}\int |\bu|^2      dx^3=\frac{1}{2}\sum_s\sum_\bk   |\tu^s_\bk|^2,\quad \mathrm{and} \quad
\cH=\frac{1}{2}\int  \bu\cdot\bw dx^3=\frac{1}{2}\sum_s\sum_\bk sk|\tu^s_\bk|^2,
\eeq 
It is worth noting that while $\cE$ is a positive quantity, the helicity can take either sign. 

\subsection{ Homochiral Navier-Stokes} 
When only the homochiral terms are kept in the NSE the system reduces to
\beq
\partial_t \bu^{s} = 
 \mathbb{P}^{s}\left[ (\bu^{s}\times \bw^{s}) \right]
+ \nu \nabla^2 \bu^{s} + \bf f^{s}. 
\label{HCNS}
\eeq
In this case the two helical fields $\bu^\pm$ evolve independently, with the non-linearity 
conserving their energy and helicity
\beq 
\cE^\pm=\frac{1}{2}\sum_\bk |\tu^\pm_\bk|^2, \qquad
\cH^\pm=\frac{1}{2}\sum_\bk \pm k|\tu^\pm_\bk|^2.
\eeq
However, unlike the Navier Stokes equation, in the homochiral version the two
helicities are sign definite quantities with $\cH^+> 0$ and $\cH^-<0$. 
The sign definiteness of the helicity  has a profound impact on the cascade. 
As shown in \cite{biferale2012inverse,biferale_musacchio_toschi_2013}, it leads to an inverse cascade. 
In fact when $\cH^\pm$ are sign definite one can show that a simultaneous 
forward cascade of $\cE^s$ and $\cH^s$ is incompatible (using similar arguments to  Fjortoft  \cite{fjortoft1953changes} 
for the dual cascade of energy and enstrophy in two dimensions, see \cite{alexakis2018cascades} sec. 3.5).
On the contrary, for the original Navier-Stokes case ($\lambda=1$), and in the presence of a large scale helical forcing, the Energy and Helicity cascades are observed to be forward as originally proposed in
\cite{brissaud1973helicity} and later verified numerically \cite{chen2003joint,alexakis2017helically}.
\subsection{The $\lambda$-Navier-Stokes model}
The different roles played by the homochiral and heterochiral interactions lead Sahoo et al \cite{sahoo2017discontinuous} to propose a model that transitions from the forward cascading Navier-Stokes \ref{HNS} to the inverse cascading homochiral Navier-Stokes \ref{HCNS} varying continuously a parameter $\lambda$.
In detail, the model reads:
\beq
\partial_t \bu^{s} = 
 \mathbb{P}^{s}\left[ (\bu^{s}\times \bw^{s}) \right]+
\lambda \mathbb{P}^{s}\left[ (\bu^{-s}\times \bw^{s}) + (\bu^{s}\times \bw^{-s}) +(\bu^{-s}\times \bw^{-s}) \right]
+ \nu \nabla^2 \bu^{s} + \bf f^{s}, 
\label{lambda}
\eeq 
For $\lambda=1$ homochiral and heterochiral terms are balanced so that one recovers the Navier-Stokes equation \ref{HNS} 
where energy cascades forward. For $\lambda=0$,  the heterochiral terms are eliminated and the system reduces to the homochiral NSE \ref{HCNS}. For any finite value of $\lambda$ the system \ref{lambda} has exactly 
the same ideal invariants as the NSE $\cE=\cE^++\cE^-$ and helicity $\cH=\cH^++\cH^-$. As for the original  NSE,
$\cH$ is not a sign definite quantity and thus it poses no restriction in the direction of the energy cascade. One can not thus
trivially predict the direction of the energy transfer when $\lambda \neq 0$. 

In \cite{sahoo2017discontinuous} it was shown that as the parameter $\lambda$ was varied from $1$ to $0$ a change of the cascade direction  was observed from a forward to an inverse cascade. In the limit of infinite Reynolds number
this transition was shown to converge to a critical discontinuous transition at a critical value $\lambda_c\simeq0.3$
such that for $\lambda<\lambda_c$ all injected energy cascades to large scales while for $\lambda>\lambda_c$ 
all energy cascades to the small scales.


\section{Numerical set-up}
In the present work we are going to investigate the limit $\lambda\to \lambda_c$ from above  $\lambda>\lambda_c$.
To do that, we perform numerical simulations of the $\lambda$-Navier-Stokes system \ref{lambda} in a triple periodic cubic domain
of size $2\pi L$. The velocity field $\bu$ is evolved using a pseudo-spectral code with 2/3 dialiasing and a second order
Runge-Kutta method for the time advancement. 
We use a uniform grid with $N$ grid points in each direction. The values of $N$ used varied from $N=128$
to $N=1024$ depending on the Reynolds number used.

In the examined range of $\lambda>\lambda_c$ the cascade is forward so we pick the forcing
to act only on Fourier modes that lie inside a sphere of radius $k_f=2/L$. The phases of the forced Fourier modes $\tbF_\bk$ are changed randomly at every time step so that the forcing is delta correlated in time and 
injects energy on average at a fixed rate denoted here by $\epsilon$, \LB{and with zero helicity injection}. Given the input parameters of our system, 
 the only other non-dimensional number, besides $\lambda$, is the  $\epsilon$-based Reynolds number:
\beq \Rep \equiv \frac{\epsilon^{1/3}L^{4/3}}{\nu}.\eeq 
Small resolution runs $N\le256$ started from random initial data and were evolved until a statistically steady state is reached where all quantities fluctuate around a mean value and the energy injection is balanced by the energy dissipation. 
Larger resolution runs $N\ge512$ started with initial conditions obtained from smaller resolution runs extrapolated to the new grid. They were then evolved until a statistically steady state is reached.

A table of the parameters of our  runs is given in table \ref{tbl}. \LB{Let us also stress that  simulations with values of $\lambda$ on the other side of the transition $\lambda < \lambda_c$ are hindered from the fact that in such a range one would need to fully resolve also the inverse energy  cascade range and this is by far too demanding for the scopes of this work.}
\begin{table*}[]
    \centering
    \begin{tabular}{|c|c| c| c| c| c| c| c| c|}
        \hline 
        $\lambda=$   & \quad 0.30 \quad   &  0.35    & \, 0.40 \,   & 0.45  & 0.50  & 0.60   & 0.80   & 1.00   \\  \hline
        $\Rep=500$  & $ 128 $  & $ 128 $  & $256$  & $\, 256 \,$ & $\, 256 \,$ & $\, 256 \,$  & $\, 256 \,$  & $\, 256 \,$  \\
        $\Rep=840$  &  $128$  & $256$  & $256$  & $256$ & $256$ & $256$  & $512$  & $512$  \\
        $\Rep=2500$ &   -     & $512$  & $512$  &   -   &   -   & $512$  & $1024$ & $1024$ \\
        $\Rep=6300$ &   -     & $1024$ & $1024$ &   -   &   -   & $1024$ &    -   &   -    \\
        \hline
    \end{tabular}
    \caption{Resolutions used for all the simulations performed. It is worth noting that values of $\lambda$ closer
    to the critical value $\lambda_c\simeq 0.3$ require less resolution for the same value of $\Rep$ but require longer computational time to converge to a statistically steady state.}
    \label{tbl}
\end{table*}

\section{Results}
\subsection{Energy balance relations}
\begin{figure}[!ht]
  \centering
  \includegraphics[width=0.68\textwidth]{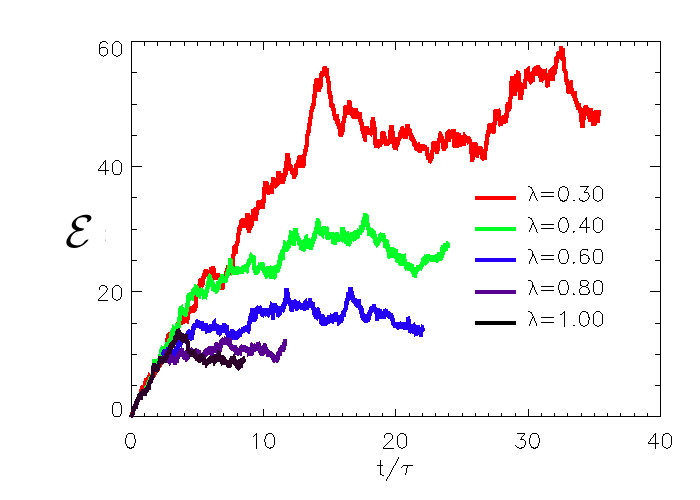}
  \caption{ Evolution of energy, $\cE$,  as a function of time for $\Rep=500$ and different values of $\lambda$. \label{fig:etime} 
  }
\end{figure}

\begin{figure}[!ht]
  \centering
  \includegraphics[width=0.48\textwidth]{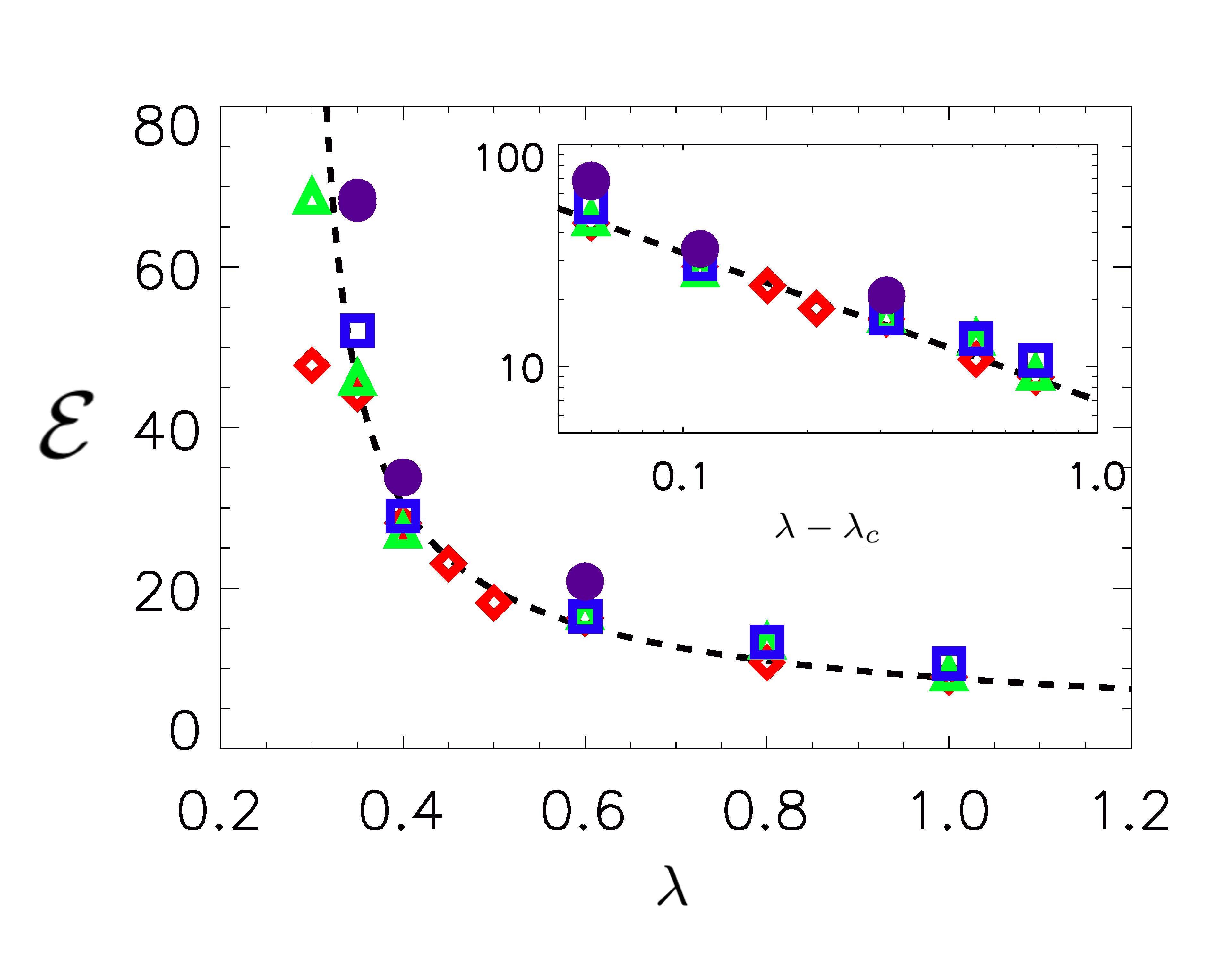}
  \includegraphics[width=0.48\textwidth]{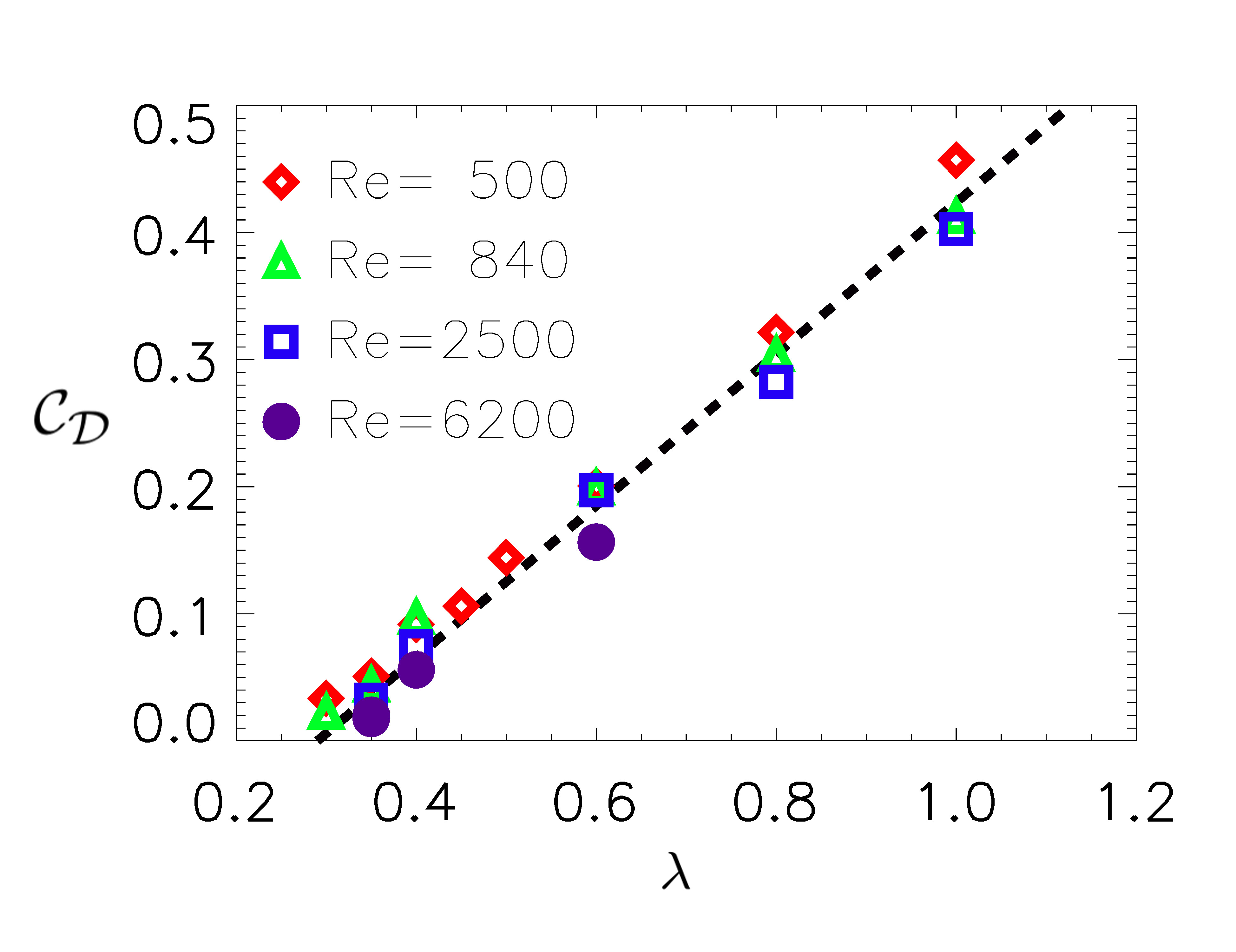}
  \caption{ Left: Energy of the flow at steady state as a function of the parameter $\lambda$. Different symbols correspond to different Reynolds numbers. The dashed lines gives the prediction \ref{scl}. \LB{The inset shows the same data as a function of $\lambda-\lambda_c$ in a log-log scale.}
  Right: Normalized energy dissipation rate $\cC_\cD$ as a function of $\lambda$. \label{fig:energy} 
  }
\end{figure}

Figure \ref{fig:etime} shows the time evolution of the total energy, $\cE$, for the smallest value of $\Rep=500$ examined and for different values of $\lambda$. As the value of $\lambda$ approaches its critical value $\lambda_c\simeq 0.3$ the mean (time averaged) energy is increased, as also are the fluctuations around the mean value. This behavior could in part be anticipated since 
by taking the limit $\lambda\to \lambda_c$ we reduce the efficiency of the flow to transport energy to the small scales
so the amplitude of the turbulent fluctuations has to increase to compensate this lack of efficiency and maintain 
the flux of energy to the small scales fixed and equal to the injection rate $\epsilon$.
In the left panel of figure \ref{fig:energy} 
we show the time averaged energy as a function of $\lambda$ for the different values of $\Rep$. 
Different symbols are used for the different values of $\Rep$ as indicated in the legend.
For large values of  $\lambda$ the amplitude of the mean energy is practically independent on the 
value of $\Rep$ (in the range examined) and is weakly dependent on $\lambda$.
However as $\lambda$ is decreased close to $\lambda_c$ the mean energy increases displaying a divergence at $\lambda_c$.
Close to $\lambda_c$ the mean energy strongly depends on the value of $\Rep$, increasing as $\Rep$ is increased.
This implies that for values of $\lambda$ close to $\lambda_c$ we have not yet reached the asymptotic state $\Rep\to\infty$
where energy saturation is independent on the value of viscosity.

An alternative way to plot the same data is to study the ratio 
\beq 
\LB{ \cC_\cD=\frac{2\pi L \,\epsilon }{U^3} }
\eeq  
where $U$ is the root mean square value
of the velocity $U=\sqrt{2\cE}$.
This ratio expresses the efficiency of turbulent fluctuations of a given amplitude to 
cascade energy to the small scales and sometimes it is called the normalised dissipation rate. 
It is a fundamental property of turbulence that 
$\cC_\cD$ remains finite in the $\Rep\to \infty$ limit, resulting in finite dissipation of energy in the zero viscosity limit.  
%
%
This quantity is plotted in the right panel of fig. \ref{fig:energy}.
\LB{For $\lambda=1$, $\cC_\cD\simeq 0.4$ that is close 
to reported values (\cite{djenidi2017normalized}) but is decreasing as $\lambda$ is decreased.}
The data indicate that it linearly approaches zero as $\lambda\to\lambda_c$, so that $\cC_\cD\propto (\lambda-\lambda_c) $. Right at the critical point $\lambda=\lambda_c$ the flow is inefficient to cascade the energy to the small scales. 
At this critical point, the flow evolution is limited only by viscous effects at the forcing scale to saturate the energy injection and the amplitude of the fluctuations would diverge in the $\nu\to0$ limit. 

The linear approach to zero can be reinterpreted to find the divergence observed in the energy  in the \LB{left} panel of \ref{fig:energy} as
\beq \cE \propto \frac{\epsilon^{2/3}}{(\lambda-\lambda_c)^{2/3}} \label{scl}.\eeq 
The dashed line in this figures shows that indeed this scaling is compatible with the data. 

There are a few comments that should follow the result in eq.\ref{scl}.
First of all, we should stress again that the increase in the energy as $\lambda\to\lambda_c$ is approached
is due a reduced efficiency of the flow to cascade energy to smaller scales.
This has some direct consequences. If we define the Reynolds number based on the rms  velocity of 
the flow  \beq \Reu \equiv \frac{UL}{\nu} \eeq 
the two definitions $\Rep,\Reu$ are not equivalent but $\Reu \propto (\lambda-\lambda_c)^{-1/3} \Rep$.

We also need to comment on the two limits $\lambda\to\lambda_c$ and $\Rep \to \infty $. 
For any value of $\lambda>\lambda_c$ the normalized dissipation rate $\cC_\cD$ 
will remain strictly positive in the $\Rep\to \infty$ limit. 
On the other hand for $\lambda=\lambda_c$, where there is no cascade to the small scales,
velocity fluctuations saturate with amplitudes such that  $\epsilon\propto\nu U^2/L^2$ leading
to the estimate $\cC_\cD\propto \Rep^{-3/2}$ which becomes zero at infinite $\Rep$. 


\subsection{Spectral properties}

Modifying the efficiency of the flow to cascade the energy to small scales will affect 
the turbulent scale-by-scale energy budget.
Further understanding can be obtained by looking at the spectral distribution of energy.
The spherically averaged energy spectrum $E(k)$ is defined as
\beq 
E(k) = \sum_{k<|\bq|<k+1} \left[ |\tu^+_\bq|^2+ |\tu^-_\bq|^2 \right]
\eeq 
and expresses the amount of energy in a  spherical shell in Fourier space with unit width. 
We note that the $\lambda$ model used here has exactly the same scaling symmetries 
as the Navier-Stokes equations. Therefore, dimensional analysis will imply again 
a Kolmogorov energy spectrum $E(k)\propto k^{-5/3}$.

\begin{figure}[!ht]
  \centering
  \includegraphics[width=0.48\textwidth]{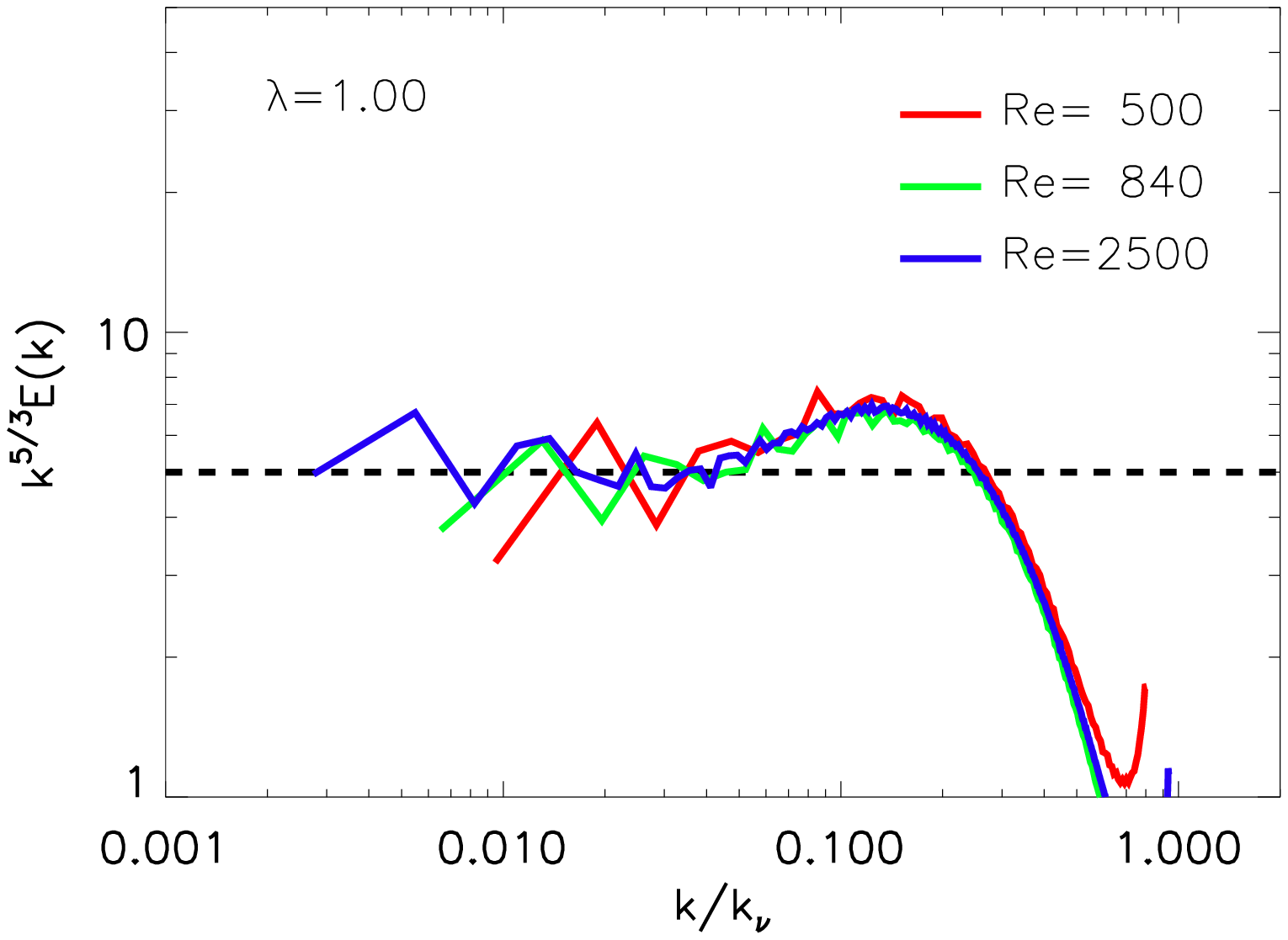}
  \includegraphics[width=0.48\textwidth]{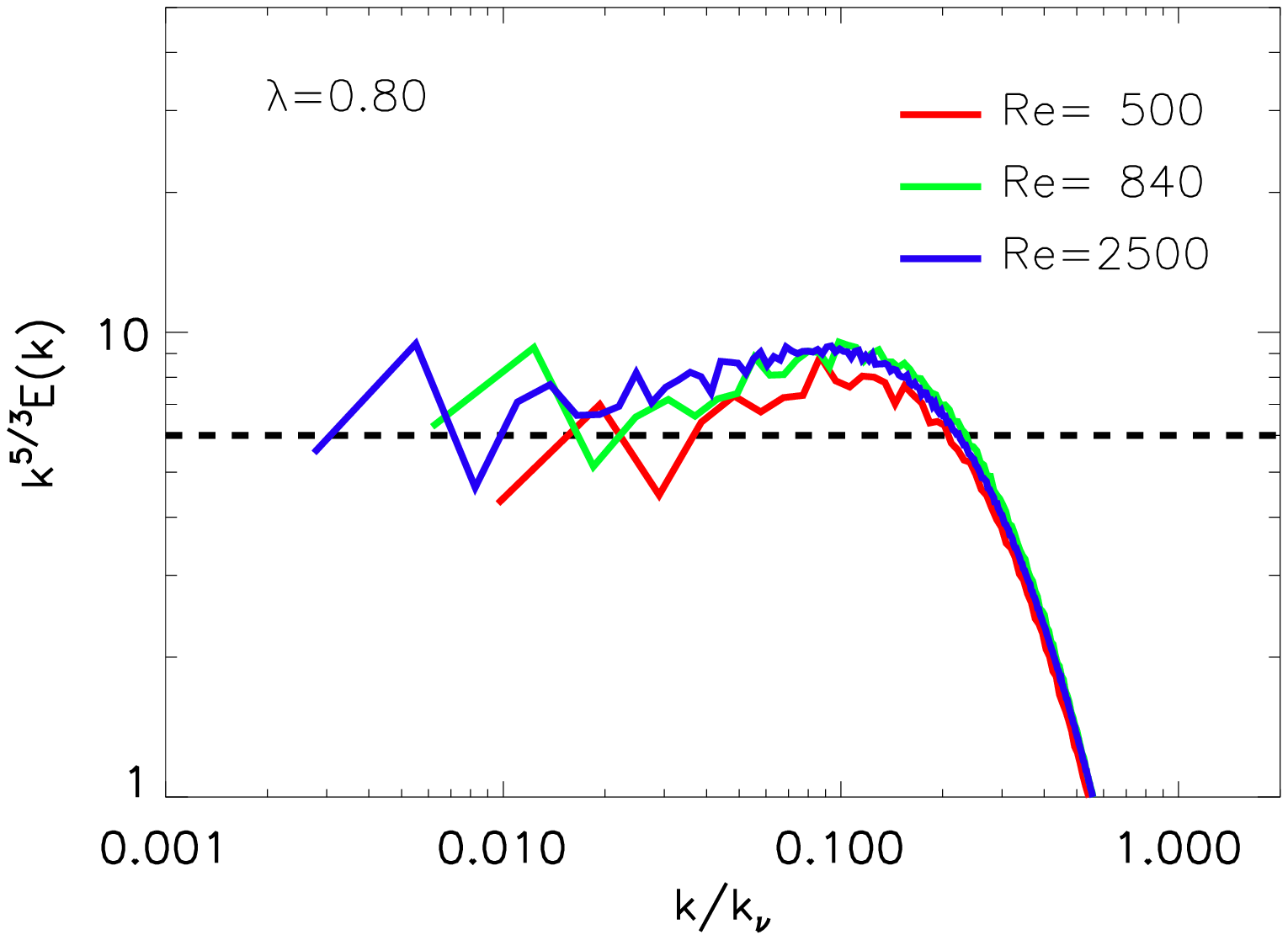}
  \includegraphics[width=0.48\textwidth]{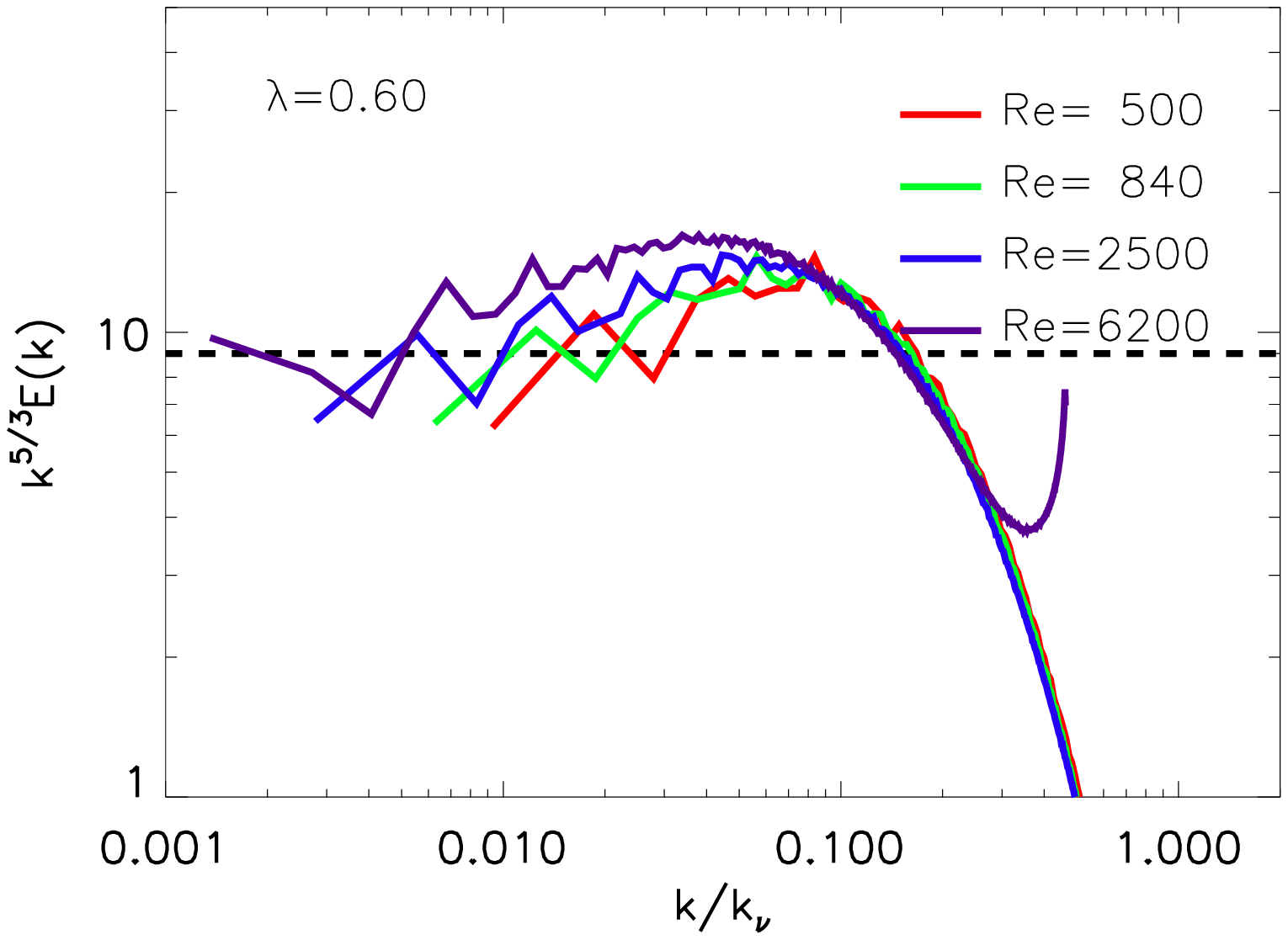}
  \includegraphics[width=0.48\textwidth]{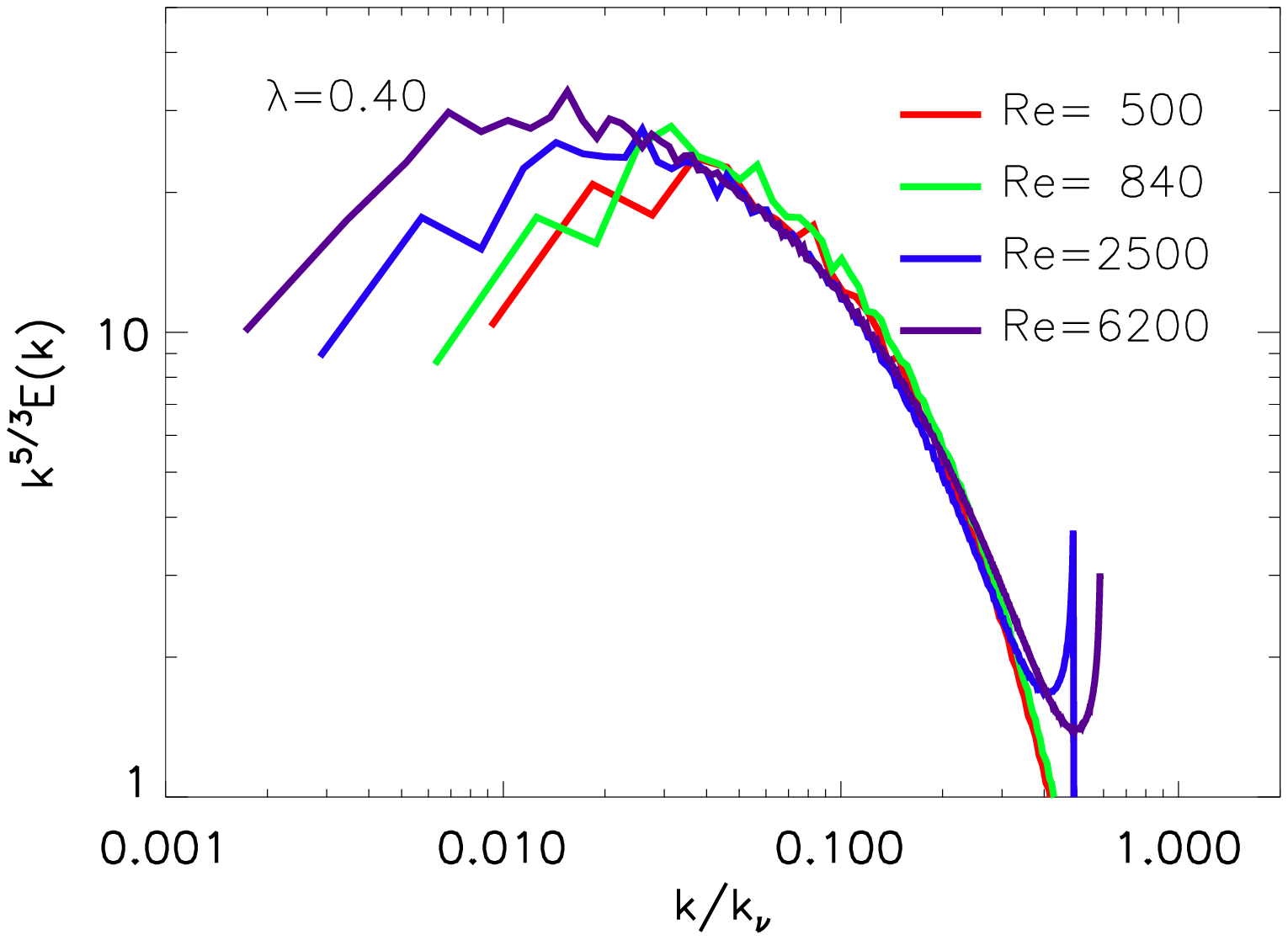}
  \caption{ Energy spectra normalized by $k^{-5/3}$ for different Reynolds numbers and for $\lambda$ 
  as indicated in the legend. Horizontal lines are put to guide the eyes and correspond to the Kolmogorov scaling. \label{fig:spectra} }
\end{figure}

The four panels of figure \ref{fig:spectra} display $E(k)$
for different values of $\Rep$ and $\lambda$. The spectra are compensated by $k^{5/3}$ so that a Kolmogorov spectrum 
 would appear as flat. The $x$-axis has been re-scaled by the Kolmogorov dissipation wavenumber
\beq 
k_\nu = \frac{\epsilon^{1/4}}{\nu^{3/4}} 
\eeq 
so that the spectra of different $\Rep$ collapse together at large wavenumbers. 

The $\lambda=1$ case shows the typical large $\Rep$ behavior of a turbulent flow
for which the power-law behavior $k^{-\alpha}$, (with $\alpha \sim -5/3$), 
is followed by a bottleneck increase of the compensated spectrum.
The bottleneck behavior is well documented in the literature (\cite{falkovich1994bottleneck,donzis2010bottleneck}) and is roughly explained 
as a pile up of the cascading energy when the viscous cut-off is reached. 
In \cite{frisch2008hyperviscosity} it was argued that when a very high order hyperviscosity 
is used the bottleneck is increased approaching a {\it thermalized} state.
Thermalized states manifest themselves in conservative systems like the spectraly truncated Euler equations 
in which energy is equally distributed among all modes leading to the 
energy spectrum $E(k)\propto k^2$ \cite{kraichnan1973helical}. The
transition of hyperviscous runs to a thermalized state has been recently demonstrated
in \cite{agrawal2020turbulent}. For regular viscosity the bottleneck has thus been interpreted
as a partial {\it thermalization}. 

\begin{figure*}[!ht]
  \centering
  \includegraphics[width=0.85\textwidth]{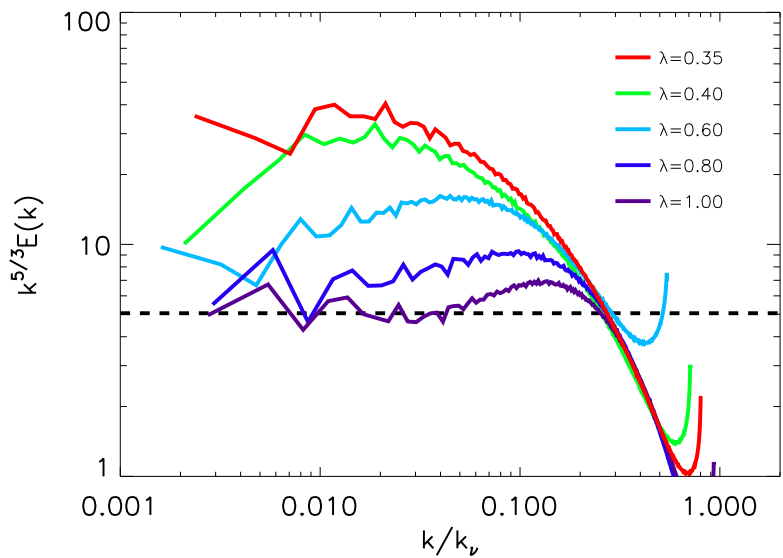}
  \caption{ Energy spectra normalized by $k^{-5/3}$ for the largest attained Reynolds numbers overlapped for the different values of $\lambda$.  Horizontal lines are put to guide the eyes and corresponds to the Kolmogorov scaling. \label{fig:spectra2}}
\end{figure*}

As $\lambda$ is decreased this behavior starts to change. The bottleneck
appears to increase in amplitude and covers a wider range of wavenumbers. 
This tendency is demonstrated in figure \ref{fig:spectra2} where the spectra are plotted 
fro different values of $\lambda$. For $\lambda=0.8$ one can still observe a $k^{-5/3}$ 
range but for smaller values of $\lambda$ it is hard to observe a power-law range
with the present resolution. At the smallest values of  $\lambda$ examined $\lambda=0.35$ 
and $\lambda=0.40$ the bottleneck covers the whole 
range of wavenumbers.
Therefore as $\lambda\to \lambda_c$ the range of wavenumbers which follow partial thermalisation increases. \LB{It is impossible, within the given resolution limitations, to precisely estimate the scaling of the extension of the bottleneck effects as a function of $\lambda$ and Reynolds number.}
An understanding of why this excess of thermalisation  occurs as the critical point is approached 
is obtained by looking at the spectral energy fluxes.

The energy flux gives the rate that energy flows across a a particular scale.
It is defined as
\beq 
\Pi(k) = -\sum_s \langle \bu^<_k \cdot \bu\cdot \nabla \bu \rangle
\eeq
where $\bu^<_k$ stands for the velocity field filtered so that only wavenumbers with norm $|\bk|<k$ are kept.
It has been shown \cite{alexakis2017helically} that can be decomposed to a homochiral part stemming 
from same chirality interactions and a heterochiral part stemming from cross-chirality interactions. 
The homochiral flux is defined as
\beq 
\Pi^{homo}(k) = -\sum_s \langle \bu^<_k \cdot \mathbb{P}^{s}\left[ (\bu^{s}\times \bw^{s}) \right] \rangle
\eeq 
while the heterochiral flux is defined as
\beq 
\Pi^{hete}(k) = -\lambda \sum_s \langle \bu^<_k \cdot \mathbb{P}^{s}\left[ (\bu^{-s}\times \bw^{s}) + (\bu^{s}\times \bw^{-s}) +(\bu^{-s}\times \bw^{-s}) \right] \rangle.
\eeq
The total flux is equal to the sum of the two 
\beq
\Pi^{hete}(k)=\Pi^{homo}(k)+\Pi^{hete}(k).
\eeq 
\begin{figure*}[!ht]
  \centering
  \includegraphics[width=0.49\textwidth]{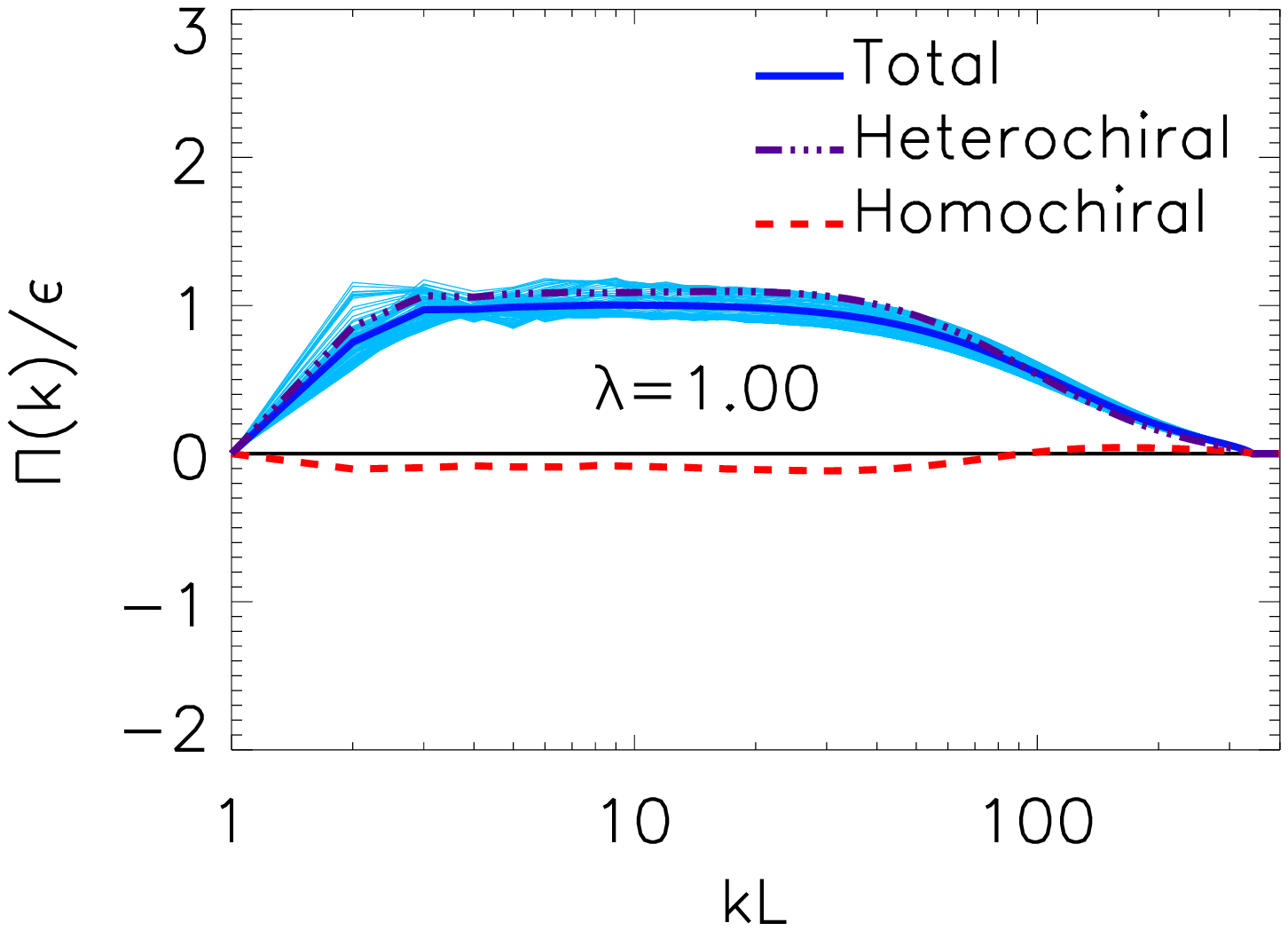} 
  \includegraphics[width=0.49\textwidth]{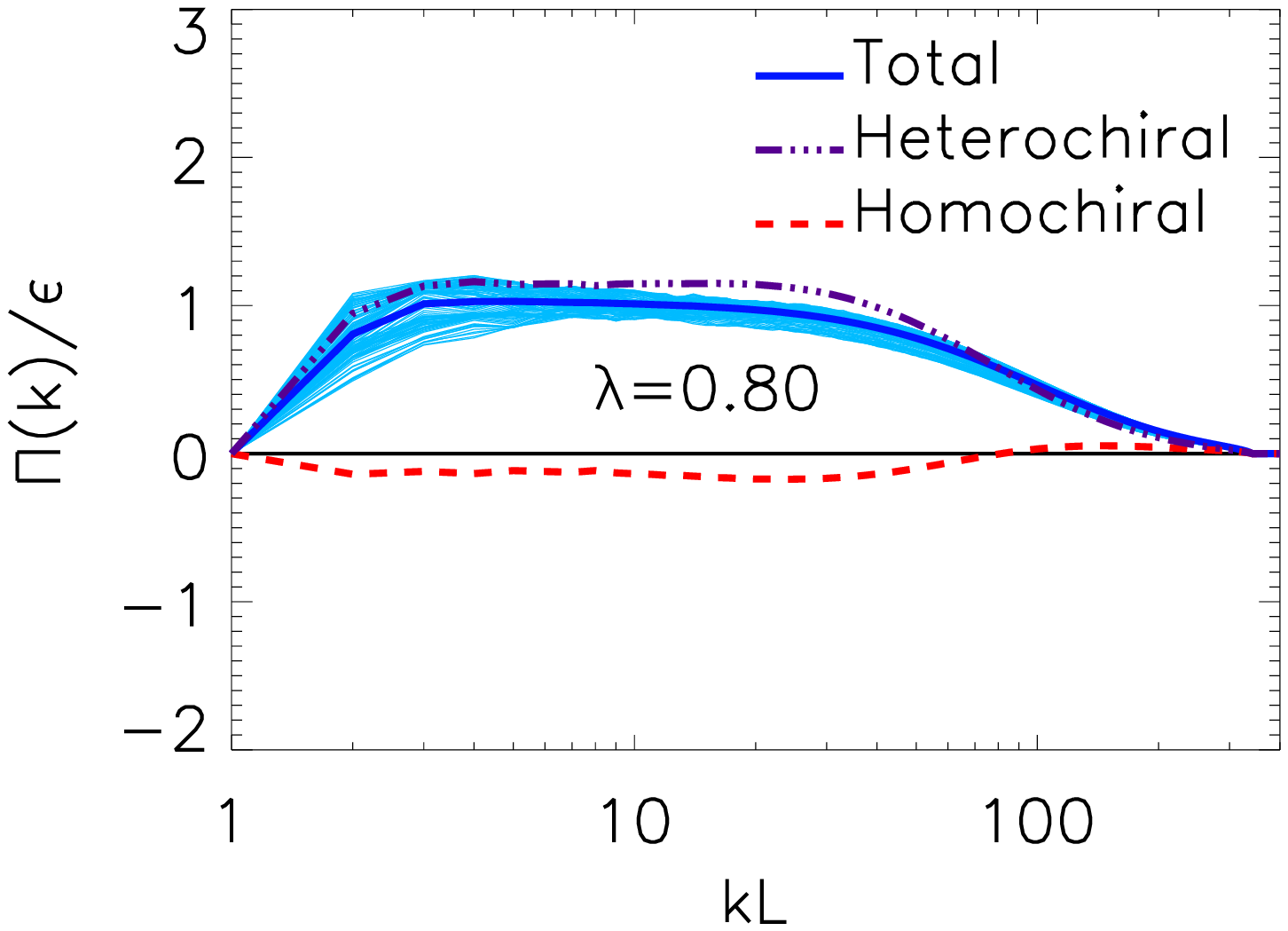} 
  \includegraphics[width=0.49\textwidth]{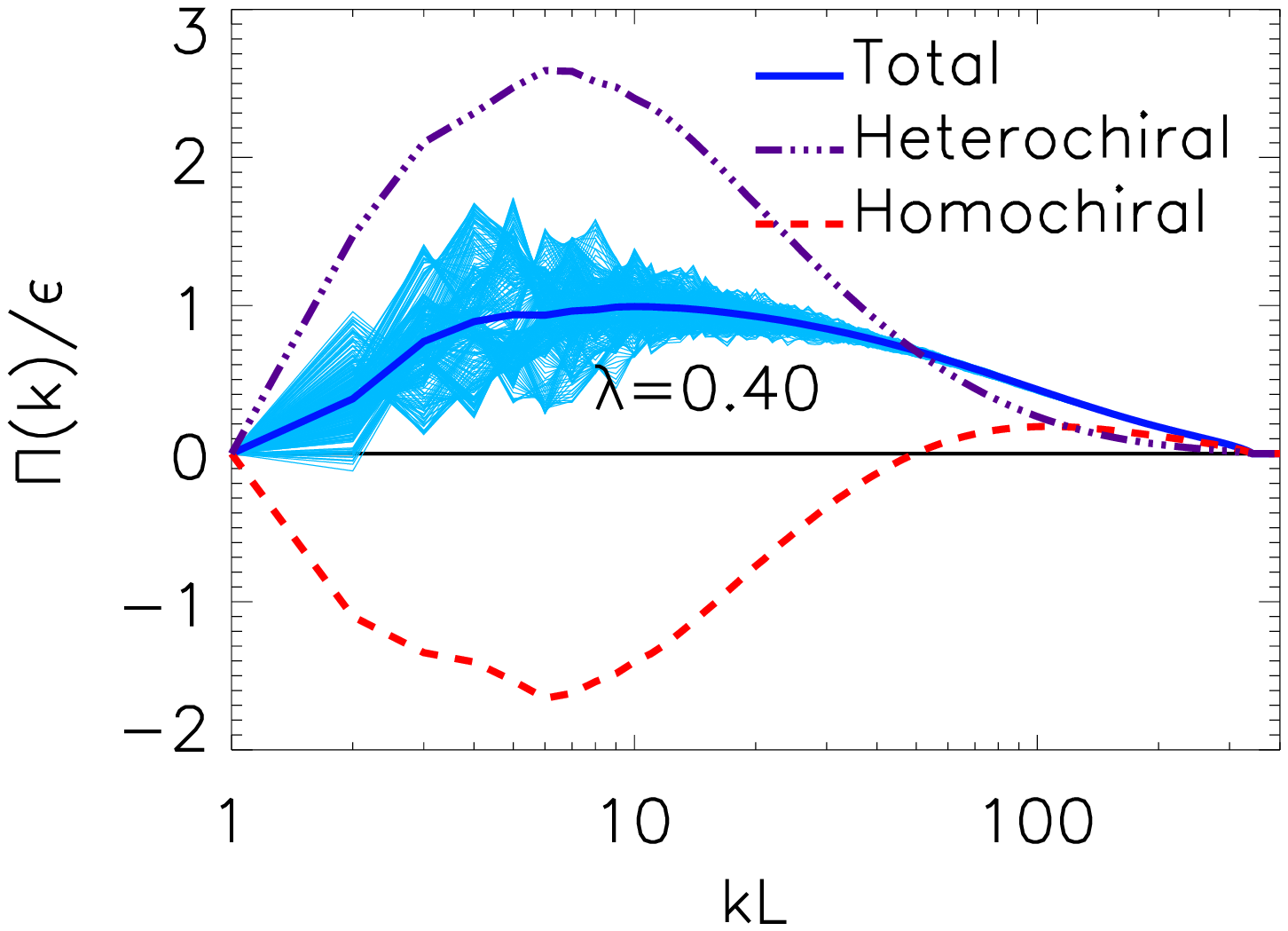} 
  \includegraphics[width=0.49\textwidth]{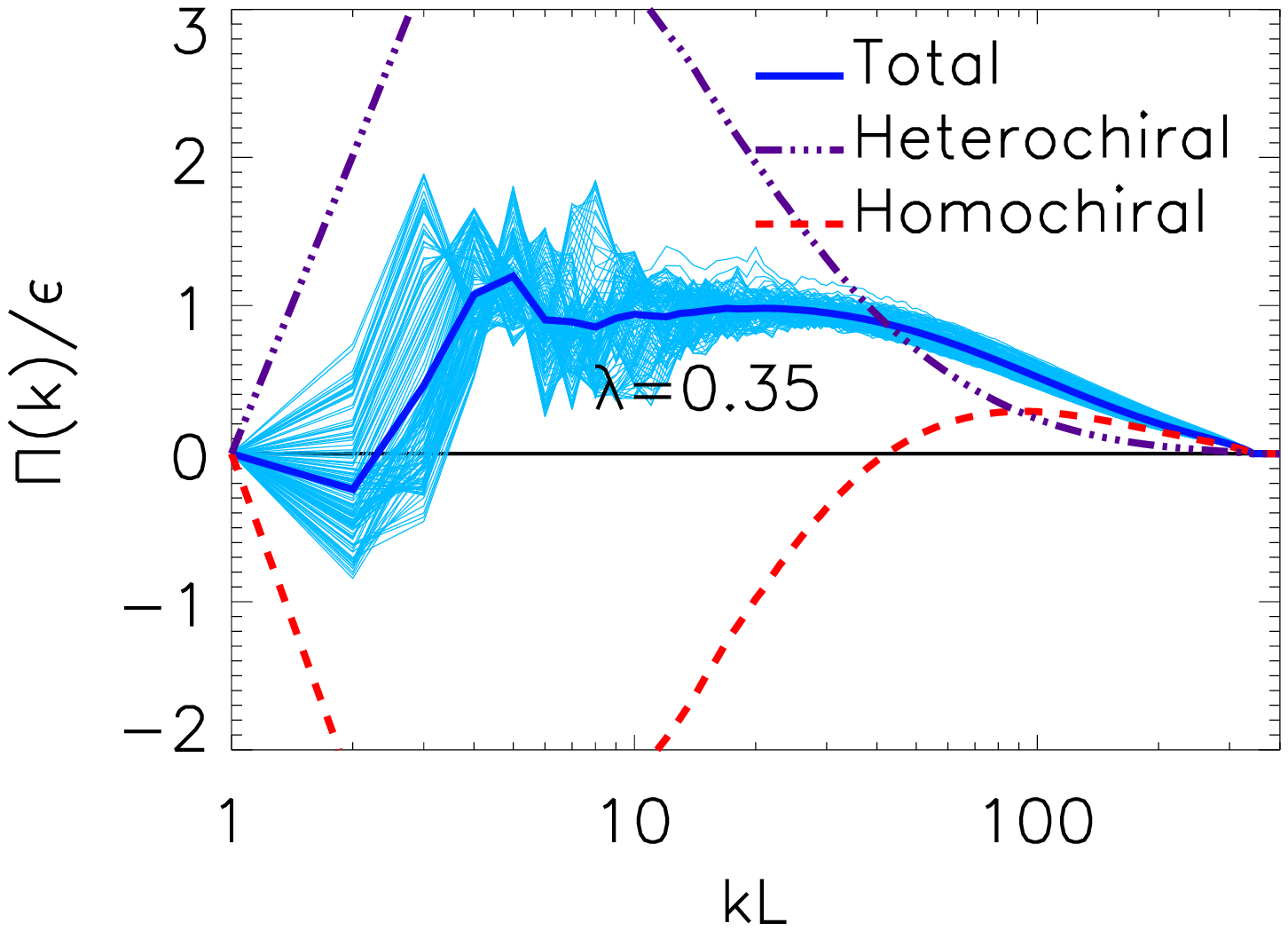} 
  \caption{Fluxes: total time averaged (blue solid line), homochiral time averaged (red dashed line), \& heterochiral time averaged (purple dash-dot line), instantaneous total fluxes (light blue)
  \label{fig:fluxes}}
\end{figure*}
In the four panels of figure \ref{fig:fluxes} we show the fluxes for four
different values of $\lambda=0.35,0.4,0.6,1.0$ for the largest resolutions attained.  
The time averaged total energy flux $\Pi(k)$ is shown with a dark blue solid line. 
It has been decomposed to its homochiral (red dashed line) and a heterochiral component
(purple dashed-dot line). With a light blue lines the instantaneous total energy flux is
shown for several different times.  
In all cases the total energy flux is positive and equal to the energy injection/dissipation rate,
while the homochiral flux is negative and the heterochiral flux is positive.  
For the Navier-Stokes case $\lambda=1$ the negative homochiral flux constitutes a small 
fraction of the total flux (about 10\%) so that $-\Pi^{homo}\ll\Pi^{hete}$.  
Small fluctuations around the time averaged value are observed in the instantaneous fluxes. 
As $\lambda$ approaches the critical value $\lambda_c$ the amplitude of negative homochiral
flux and the positive heterochiral flux both increase, keeping of course their sum 
fixed to the injection rate. As a result, the two competing processes for the 
transfer of energy to smaller and larger scales come closer together in amplitude
making their relative difference smaller and smaller. This leads to also 
an increase in the amplitude of the fluctuations around the mean value 
observed. On one hand,  the increase of fluctuations with respect to the mean could be a  potential indication that the system is driven toward a sort of quasi-thermal state, that could interpret the large bottleneck as partial thermalization as it \cite{frisch2008hyperviscosity}. On the other hand, the existence of two counter-directional fluxes one driving energy toward small scales and the other oppositely with clear non-zero average and amplitudes that becomes larger and larger by approaching $\lambda_c$ is an indication that the fluid remains  out of equilibrium but in a flux-loop state  where finite fluxes exist that however cancel each other \cite{alexakis2018cascades}.

\subsection{Intermittency}

\begin{figure*}[!ht]
  \includegraphics[width=0.45\textwidth]{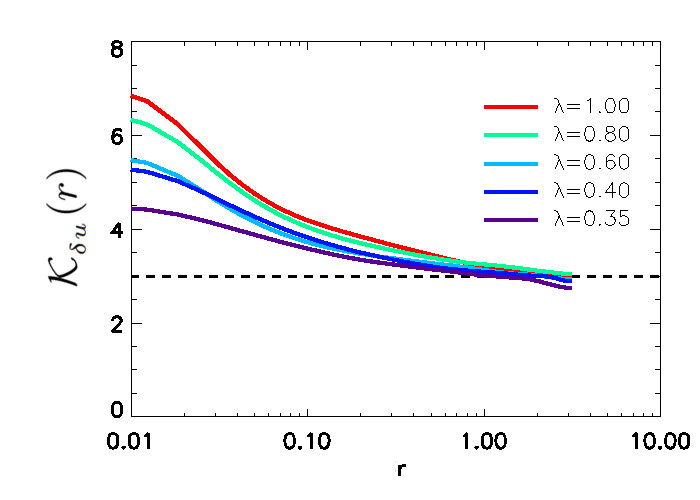}
  \includegraphics[width=0.45\textwidth]{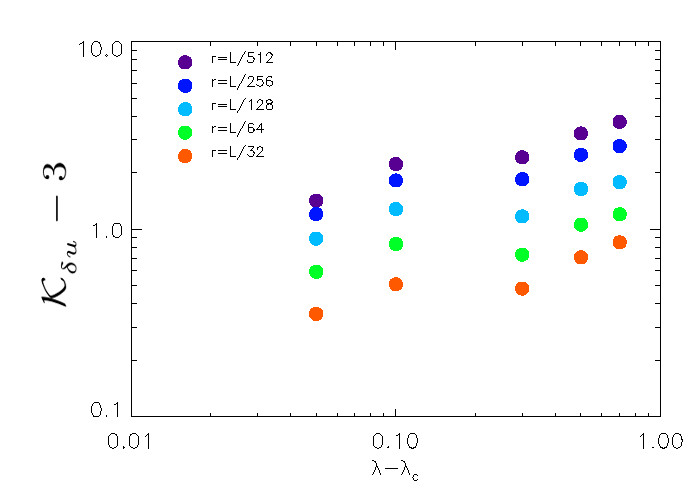}
\caption{Left: 
The Kurtosis $\mathcal{K}_{_{\delta u}}$ of the parallel velocity difference as a function of $r$
for different values of $\lambda$. 
Right: $\mathcal{K}_{_{\delta u}}-3$ as a function of $\lambda-\lambda_c$ for different values of $r$
in a log log scale. Data
are obtained from the highest $\Rep$ runs. \label{fig:KurtosisU}} 
\end{figure*}

A key property of Navier-Stokes turbulent cascade is the presence of intermittency manifesting itself as an breaking of scale-similarity, with stronger ``{\it events}" appearing as smaller scales are examined. Such deviations from scale similarity are measured by examining the scaling behavior of  structure functions of different order
$ S_n(r) = \left\langle (\delta u_r)^n\right\rangle $ where $\delta u_r$ stands for either the longitudinal increment, i.e. when ${\bf r}$ is parallel to  $\bu (\bx,t) - \bu (\bx+{\bf r},t) $ or for the transverse,  when ${\bf r}$ it is perpendicular to the velocity increment, and the brackets 
stand for a space and time average. It is well known that in Homogeneous and Isotropic Navier-Stokes three-dimensional turbulence, structure functions enjoys anomalous scaling, $S_n(r) \sim r^{\zeta_n}$, with power-law behaviours and exponents that depart from the Kolmogorov mean-field prediction, $\zeta_n \neq n/3$ \cite{frisch2012turbulence}. This is a signature of intermittency, i.e. that normalized and standardized probability distribution functions (PDF) of velocity increments cannot be superposed at changing the distance $r$ and develop a stronger and stronger departure from Gausssian statistics.  As a result, by simultaneously decreasing the scale and increasing Reynolds one can obtain turbulent states that are 
further and further away from 
a quasi-equilibrium  distribution \cite{frisch2012turbulence}. 
\LB{In our simulations, the limitation on the numerical resolution, imposed by the need to perform many different investigations for different $\lambda$ values, and the appearance of a strong bottleneck by approaching $\lambda_c$ result in the absence of a well developed  scaling range. As a result, we refrain from giving any quantitative measurements on the scaling exponents $\zeta_n$. On the other hand, in figure \ref{fig:KurtosisU} we show a dimensionless measurement of the departure from Gaussianity by plotting the Kurtosis of the velocity increments PDF at changing $r$ and for different $\lambda$ for the highest Reynolds number: }
%
\beq
\mathcal{K}_{_{\delta u}}(r) \equiv \frac{ \left\langle (\delta u_r)^4\right\rangle }{ \left\langle (\delta u_r)^2\right\rangle^2 }.
\eeq 
As one can see in the left panel,  the NSE case for $\lambda=1$ shows the classical behavior
of $\mathcal{K}_{_{\delta u}}$ increasing by decreasing $r$, going form the Gaussian value $\sim 3$ for $r \sim L$ to the highly non-Gaussian and intermittent plateau at $\sim 7$ inside the viscous range, $r \to 0$ . On the other range by decreasing $\lambda$ and approaching $\lambda_c$ we have a strong reduction in the inertial range values and also a corresponding reduction for the viscous plateau, indicating that the flow is approaching a closer and closer Gaussian distribution at all scales when $\lambda \to \lambda_c$. Similarly, on the right panel of the same figure we re-plot the data by showing the values of the Flatness at changing the distance $r/L$ and for different $\lambda$. From the latter plot there is a clearer  tendency toward the mean-field Gaussian value, $\sim 3$, by approaching $\lambda_c$ even though the behaviour is not extremely well developed. Higher Reynolds numbers are probably needed in order to enhance the critical behaviour. 

\begin{figure*}[!ht]
\centering
  \includegraphics[width=0.85\textwidth]{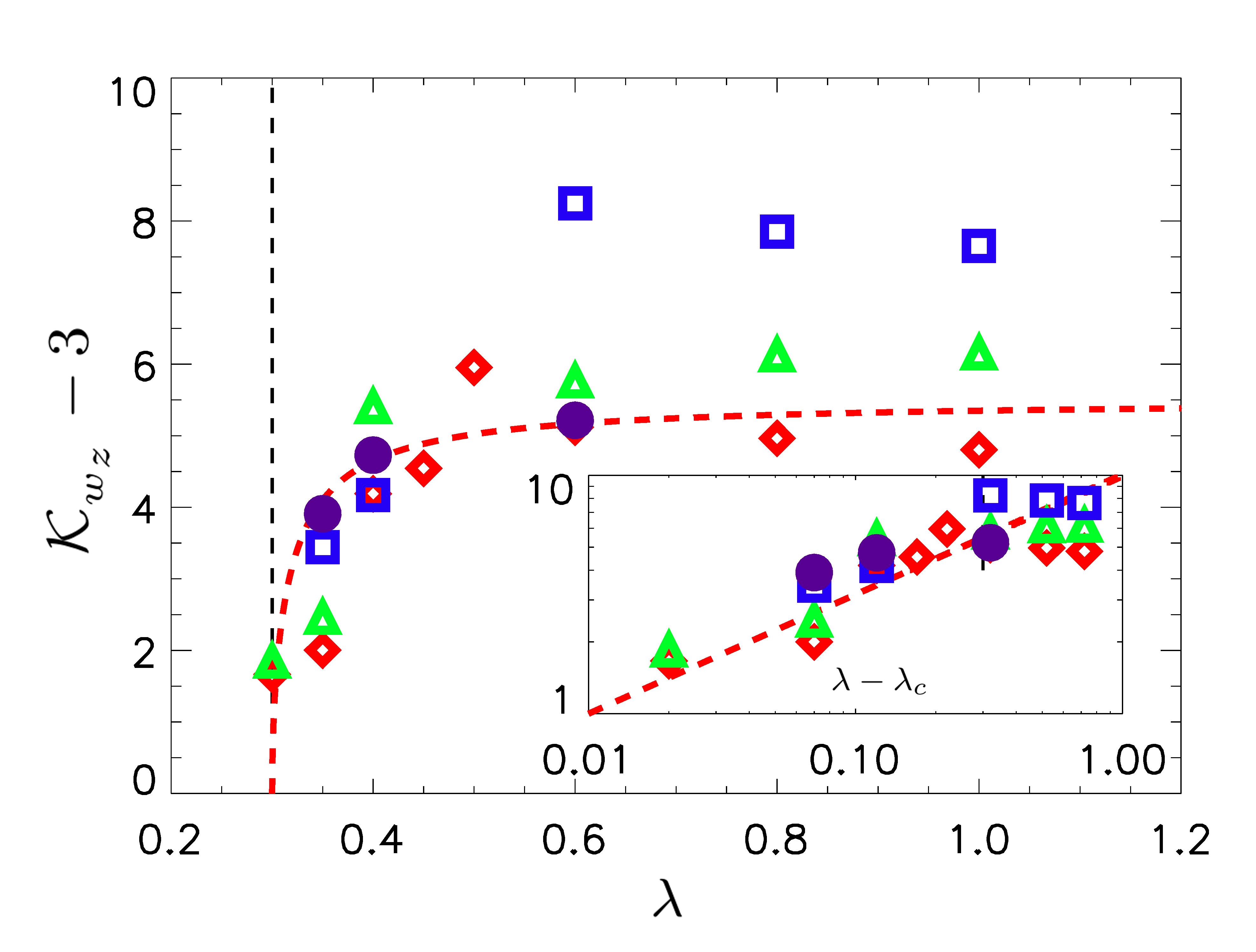}
\caption{ \LB{ Kurtosis of the vorticity PDF $(\mathcal{K}_{_{w_z}}-3)$ as a function of the different values of $\lambda$. The inset shows $\mathcal{K}_{_{w_z}}-3$ as a function of $\lambda-\lambda_c$ in a log-log scale.}
Same symbols are used as in \ref{fig:energy}: \LB{ red diamonds $\Rep=500$, green triangles $\Rep=840$, blue squares $\Rep=2500$, purple discs $\Rep=6200$. }
\label{fig:KurtosisW}  }
\end{figure*} 
Finally, because intermittency depends also on $\Rep$ we verify that the decrease of intermittency
observed in the previously examined figures is a trend that persists for all $\Rep$ examined. In figure \ref{fig:KurtosisW}
we plot the Kurtosis of the $z$ vorticity component 
\beq
\mathcal{K}_{_{w_z}} \equiv \frac{ \left\langle w_z^4\right\rangle }{ \left\langle w_z^2\right\rangle^2 }
\eeq
as a function of $\lambda$ for all examined runs.  The symbols for the different $\Rep$ used are the same as in figure \ref{fig:energy}. All data indicate that as $\lambda_c$ is approached intermittency tends to decrease. In particular, for large values of $\lambda$ the Kurtosis is increased as $\Rep$ is increased, something well known in turbulence theory due to intermittency \cite{nelkin1990multifractal,benzi1991multifractality} while as $\lambda_c$ is approached, the values appear to weakly depend on $\Rep$ and tend to the Gaussian value,  $\to 3$ for all $\Rep$.

\section{Conclusions}  

Numerical results for a variant of the Navier-Stokes equations have been presented. 
The model is characterized by a dimensionless control parameter $\lambda$  that weights homochiral and  heterochiral interactions such as to link in a continuous way the forward cascading Navier-Stokes case for $\lambda=1$ and the inverse cascading homochiral Navier-Stokes for $\lambda=0$. 
In the investigation both $\lambda$ and the Reynolds number $\Rep$ were varied in order to investigate the behavior of the system for the different values of $\lambda$ as $\Rep\to\infty$.

We have focused on the transition from the $\lambda=1$ case to case
observed at $\lambda_c \sim 0.3$ where the direct energy transfer stops and a highly complex statistical state develops.
{ 
We have shown that by approaching $\lambda_c$ from above we have a tendency to develop a more and more intense spectral viscous bottleneck. We should also note here that the limits $\lambda\to\lambda_c$ and $\Rep\to\infty$
do not necessarily commute. I.e., when the $\lambda$ limit is taken first 
the bottleneck would occupy all scales, while when  $\Rep \to \infty$ at fixed $\lambda$  a power-law inertial range is present.}
Furthermore, in the $\lambda\to\lambda_c$ limit larger and large fluctuations around the mean energy flux exist whose energy is diverging like $(\lambda-\lambda_c)^{2/3}$, accompanied by  larger and larger heterochiral and homochiral opposite contributions to the energy cascade. 
{ 
As a result, we are closer and closer to a turbulent state
where the finite counter-directing fluxes cancel each other leading to a subdominant mean energy flux. We refer to this state as flux-loop. 
Such flux-loop states have been met before in the literature in different contexts 
\cite{boffetta2011flux,di2020phase,falkovich2017vortices,musacchio2019condensate,van2019condensates} but are inadequately explored, and it is not 
known if methods from equilibrium dynamics could be applied to them.}

\LB{An other direction that could be followed would be the use of renormalization group techniques for the $\lambda$-Navier-Stokes system in $d$ dimensions.
This could lead to a $\lambda$-dependence of the 
renormalized transport coefficients that could change sign or the appearance of new $\lambda$-dependent terms.
Ideally an optimal path in the $(\lambda,d)$ could
exist where the $d=3$, $\lambda=1$ case could be solved for.}

{ 
In the present investigation, the statistics is observed to 
have a decreasing Kurtosis for both velocity increments in the inertial range and vorticity components as the flux loop state is approached,  $\lambda\to\lambda_c$.  
The results thus indicate the possibility that the flux-loop state 
could have Gaussian or quasi-Gaussian statistics.}
The entangled presence of non-zero fluxes and reduction of non-Gaussian contributions opens the way to perturbative approaches of intermittency, offering also a unique testing bed for any new theory of turbulence because of the possibility to change the Navier-Stokes statistics as a function of a free control parameter.

\vskip 6pt

\enlargethispage{20pt}



\aucontribute{All authors participated in the analytical computations. AvK drafted the manuscript and performed the numerical simulations. All authors read, edited and approved the manuscript.}

\competing{The authors declare that they have no competing interests.}

\funding{This  work  was  supported  by  Agence  nationale  de  la  recherche  (ANR  DYSTURB  project  No. ANR-17-CE30-0004).  This project has received funding also from the European Research Council (ERC) under the European Union’s Horizon 2020 research and innovation programme (grant agreement No 882340). }

\ack{LB acknowledges useful discussions with Roberto Benzi. This work was granted access to HPC resources of MesoPSL financed by Region Ile de France and the project Equip@Meso (reference ANR-10-EQPX-29-01) of the  programme  Investissements  d’Avenir  supervised  by  Agence  Nationale  pour  la  Recherche  and  HPC resources of GENCI-TGCC \& GENCI-CINES (Projects No.  A0080511423,  A0090506421).  }


\bibliography{lambda} 
\bibliographystyle{ieeetr}
\end{document}